\documentclass{article}
\newcommand{\be}{\begin{equation}}
\newcommand{\ee}{\end{equation}}
\newcommand{\bea}{\begin{eqnarray}}
\newcommand{\eea}{\end{eqnarray}}
\newcommand{\nn}{\nonumber}
\usepackage{amsmath}
\usepackage{amsfonts}
\usepackage{amssymb}
\begin{document}

%
%

{\center Quantum Properties of Periodic Instantons on a Circle \\
\vspace{10mm}

A. V. SHURGAIA\footnote{e-mail: avsh@rmi.acnet.ge} \\
\vspace{5mm}
Department of Theoretical Physics \\
A. Razmadze Mathematical Institute  \\
  1 M. Alexidze str. 0193 Tbilisi \\
             Georgia \\}




\begin{abstract}
Quantum properties of a (1+1)-dimensional scalar theory on a cylinder with a compact spatial part, namely, $0\leq x\leq L$, are considered. In particular,  quantum theory around the classical periodic field configurations is studied and the lifetime of a quantum periodic instanton is estimated.\\

\end{abstract}
\vspace{15mm}

1. Periodic field configurations known as the periodic instantons have been for many years  in the field of view of researchers in
view of their important role in the process of quantum tunneling and the associated phase transitions. The attractiveness of such
field objects is that, depending on the total energy they interpolate between the vacuum and saddle point states (sphalerons) and
thus have the ability to be responsible for the phase transition from the classical processes at high energies to quantum tunneling
ones at low energies. The complexity of actual physical theories does not allow to obtain exact results, hence lower dimensional
models which make  it possible to perform some analytical calculations are of interest. From this point of view (1+1)-dimensional
scalar theory is subjected to intensive study. Along with stable field configurations of finite energy more recently unstable
configurations like bounces\cite{cole}, sphalerons\cite{man,klin}, periodic instantons\cite{sam,muel} have been investigated. Recently, interest has grown in configurations on compact space and configurations like (anti)periodic (called also twisted fields) are in addition investigated\cite{sak}-\cite{paw2}. In the cited papers the exact solutions of classical equations of motion on a circle, as well as the equation  of fluctuations  are investigated and the  regularization of the energy in one-loop approximation is carried out. In the present paper we are interested in periodic classical field configurations and their quantum properties in two-dimensional scalar theory, the spatial  part of which is a circle with circumference $L$. There is an interesting interpretation of nontrivial classical filed configurations for finite $L$, provided, that the potential of the model we explore has two distinct vacua separated by a finite potential barrier. This allows to consider the process of creating of kink-antikink pair at some energy. They propagate along the circle in opposite directions and annihilate at the meeting points leaving the field in the other vacuum. The largest distance between kink and antikink corresponds to the state at the top of a potential barrier - that is the saddle point of the energy functional.  With  increasing $L$ unstable configurations of many kink-antikink states can be appeared\cite{sam}. In the present paper we start from the quantum theory in Schr\"{o}dinger representation. We use the symmetry properties of the theory and construct a perturbation theory in the inverse powers of a coupling constant using the collective coordinates method (see for general theory and some application\cite{shur} and references therein). We calculate the quantum energy levels and estimate the lifetime of the system.

2.  Thus, we consider the system with the lagrangian density
\bea
\mathcal{L}=\frac{1}{2}\left(\frac{\partial\phi(t,x)}{\partial x_\mu}\right)^2-U(\phi(t,x),g)
\eea
with the potential satisfying
\bea
U(\phi(t,x),g)=\frac{1}{g^2}U(g\phi(t,x),1) \nn
\eea
in which $0\leq x\leq L$.
We work in the Schr\"{o}dinger representation and look for solutions of the equation:
\bea
(H-E)\Psi(\phi(x),\pi(x))=0.
\eea
The corresponding quantum Hamiltonian is
\bea
H=\int_0^Ldx\left\{\frac{1}{2}\pi(x)^2+\frac{1}{2}\left(\frac{\partial\phi(x)}{\partial x}\right)^2+U(\phi(x),g)\right\}
\eea
subject to the following commutation relation:
\bea
[\phi(x),\pi(y)]=i\delta (x-y).
\eea
We now develop a perturbation  theory  for the double-well potential  $U(\phi(x),g)$:
\bea
U(\phi(x),g)=\frac{\mu^2}{2g^2}(\phi(x)^2-g^2)^2.
\eea
The system possesses a translational invariance and corresponding   conserved quantity is the momentum of the system:
\bea
P=-\int_0^Ldx\frac{\partial\phi(x)}{\partial x}\pi(x)
\eea
Let $\phi_0(x)$ be a c-number and consider the following  transformation of the field $\phi(x)$:
\bea
\phi(x)=g\phi_0(x-q)+\Phi(x-q).
\eea
The parameter $q$, being called a collective coordinate, together with the new field $\Phi(x)$ form now  a new set of variables (operators) of the system: $\{\phi(x),\pi(x)\}\rightarrow\{q, p_q, \Phi(x),{\widetilde{\Pi}(x)}\}$. Thereby the phase space has been expanded and in order to keep the number of independent variables unchanged the additional condition must be imposed  (actually this is a gauge condition):
\bea
\int_0^LdxN(x)\Phi(x)=0,
\eea
in which $N(x)$ is an arbitrary function. Really we need one more additional condition, but as we will see below this condition will appear as a constraint in the extended space. One can always choose $N(x)$  such that the relation
\bea
\int_0^LdxN(x)\frac{\partial\phi_0(x)}{\partial x}=1.
\eea
holds. The operator $\pi(x)$ is considered as a functional derivative  $ -i\delta/\delta\phi(x) $ in a Hilbert space spanned by the vectors $|\Psi>$ and hence should be expressed through the set of new variables. For this we need to define the projection operator $A(x,y)=\delta (x-y)-\frac{\partial\phi_0(x)}{\partial x}N(y)$, with the properties
\bea
\int_0^LdyA(x,y)\frac{\partial\phi_0(y)}{\partial x}=\int_0^LdxN(x)A(x,y)=0. \nn
\eea
 Now it is easy to show that the operator $\pi(x)$ can be rewritten as follows:
\bea
\pi(x)=\Pi(x-q)-\frac{N(x-q)}{g(1+\frac{1}{g}F)}\left\{p_q+M(\Phi,\Pi)\right\},
\eea
in which the quantities $F$ and $M(\Phi ,\Pi)$ are defined by the following equalities:
\bea
F=\int_0^Ldx N(x)\frac{\partial\Phi(x)}{\partial x}, \quad  M(\Phi,\Pi)=\int_0^Ldx\frac{\partial\Phi(x)}{\partial x}\Pi(x).
\eea
Besides the operator $\Pi(x)$, determined as
\bea
\Pi(x)=\int_0^LdyA(y,x)\frac{\delta}{i\delta\Phi(y)},
\eea
satisfies the condition
\bea
\int_0^Ldx\frac{\partial\phi_0(x)}{\partial x}\Pi(x)=0,
\eea
that is the  constraint of theory (that we have mentioned above) being considered as a system with constraints. The operators $q$ and $p_q=\frac{\partial}{i\partial q}$ are conjugate to each other such that $[q,p_q]$=i. Thus all extra degrees of freedom are now fixed. The operator $\Pi(x)$ obeys the following commutation relation:
\bea
[\Phi(x),\Pi(y)]=iA(x,y).
\eea
One can now check by substituting (10) into (6) that the momentum operator of the system coincides with $p_q$  and consequently $P=p_q$.
On substituting (7) and (10) into (3) one obtains the Hamiltonian in the new phase space. We see, that $\pi(x)$ depends on $q$. In order to obtain the Hamiltonian we have to integrate over $x$ in the range from $0$  to $L$. Shifting the integration variable $x$ changes the limits of integration from $-q$ to $L-q$. Nevertheless there is no explicit dependence of the Hamiltonian from $q$. Taking into consideration the periodicity of the phase space variables and the Leibnitz integral rule, it is easy to verify that $\frac{dH}{dq}$=0. This allows us to extract the $q$-dependence of the wave function of the system $\Psi(\Phi(x),q)$. Before doing so we need to transform the wave function, indeed
\bea
\Psi(\Phi,q)=e^{ig\int_0^Ldxs(x)\Phi(x)}\Psi'(\Phi(x),q).
\eea
which means that the momentum operator  $\Pi(x)$ has to be replaced by $gs(x)+\Phi(x)$. The c-number $s(x)$ is subject to the same constraint as the operator $\Pi(x)$, namely $\int_0^Ldx\frac{\partial\phi_0(x)}{\partial x}s(x)=0$. If this condition is not being fulfilled then one defines with the help of the projection operator a new quantity $s'(x)$ that obeys it. Thus the equality (10) may be rewritten as follows:
\bea
\pi(x)=\Pi(x-q)-\frac{N(x-q)}{g(1+\frac{1}{g}F)}\left\{p_q+gM(\Phi,s)+M(\Phi,\Pi)\right\}.
\eea

3. After all these one can start to solve the Schr\"odinger equation $(H-E)\Psi'(\Phi(x),q)=0$. As we have seen above the Hamiltonian $H$ does not explicitly depend on $q$ and therefore one may factorize the wave function $\Psi'(\Phi(x),q)$ as follows:
\bea
\Psi'(\Phi(x),q)=e^{ig^2Iq}\Psi''(\Phi(x)).
\eea
Thus the operator $p_q$ should be replaced by the c-number $g^2I$. We will solve the Schr\"odinger equation  by using a perturbation theory and consequently we need  now to expand the Hamiltonian in series in inverse powers of $g$. Energy and the wave function are also to expand in appropriate series:
\bea
&H&=g^2H_0+gH_1+H_2+g^{-1}H_3+...,\\
&E&=g^2E_0+gE_1+E_2+g^{-1}E_3+...,\\
&\Psi&''(\Phi(x))=\Psi_0+g^{-1}\Psi_1+... .
\eea
So we should solve the system of equations:
\bea
&&(H_0-E_0)\Psi_0=0, \\
&&(H_0-E_0)\Psi_1+(H_1-E_1)\Psi_0=0, \\
&&((H_0-E_0)\Psi_2+(H_1-E_1)\Psi_1+(H_2+E_2)\Psi_0=0.\\
&&\ldots \ldots \ldots \nn
\eea
Leading term of the Hamiltonian does not contain field operators and therefore the equation in corresponding approach  $(H_0-E_0)\Psi_0=0$ is obeyed  identically  if
\bea
E_0=H_0=\int_0^Ldx\left\{\frac{1}{2}\left(\frac{\partial\phi_0(x)}{\partial x}\right)^2+U(\phi(x),1)+\frac{1}{2}(s(x)-IN(x))^2\right\}.
\eea
The next approximation  of the system of equations we consider is:
\bea
(\Psi_0,(H_1-E_1)\Psi_0)=0,
\eea
which is linear in field operators. The regularity of the function $\Psi_0$ requires $H_1$
and $E_1$ to be identically zero.This is accomplished by using the additional conditions imposed on the field operators, and if required
\bea
s(x)-IN(x)=-v\frac{\partial \phi_0(x)}{\partial x},
\eea
and
\bea
-(1-v^2)\frac{\partial^2\phi_0(x)}{\partial x^2}+U'(\phi_0(x),1)=0
\eea
This is a pure classical equation of motion, the solution of which under periodic boundary condition for  the potential $U$ of interest is:
\bea
 \phi_0(x)=\sqrt{\frac{2k^2}{1+k^2}}{\rm \bf {sn}}(\sqrt{\frac{2}{1+k^2}}\frac{\mu x}{\sqrt{1-v^2}},k), \\
\eea
in which $0\leq k\leq 1$ is the modulus of elliptic integrals. The periodicity condition $\phi_0(x+L)=\phi_0(x)$  implies that
\bea
 L=4n{\rm \bf K}(k)\sqrt{\frac{1+k^2}{2}}\frac{\sqrt{1-v^2}}{\mu}, \\
\eea
with n integer.
So, there are critical values of the quantity $L$, at which there are bifurcations of the stationary points of the energy $E_0$.
These solutions interpolate between the stable field configurations as $k^2$
tends to $1$ (with $L$ approaching the $\infty$) and the  unstable ones (at the top of the potential barrier) for $k^2\rightarrow 0$, latter being called sphalerons\cite{sam,muel}).

    Let us now turn to the zero energy $E_0$, that takes the following form:
\bea
E_0=\int_0^Ldx\left\{\frac{1}{2}(1+v^2)\left(\frac{\partial\phi_0(x)}{\partial x}\right)^2+U(\phi(x),1)\right\}
\eea
and is for (5):
\bea
E_0=\frac{n\sqrt{2}\mu \{8(1+k^2){\rm \bf  E}(k)-[(1-k^2)(5+3k^2)+3v^2(1-k^2)]{\rm \bf K}(k)\}} {3(1+k)^{3/2}\sqrt{1-v^2}},
\eea
which for v = 0 is consistent with the expression obtained in\cite{sam}. Using the equality  (26) and the condition that is imposed on $s(x)$
one can obtain for the momentum $I$:
\bea
I=\frac{8n\sqrt{2}\mu [(1+k^2){\rm \bf  E}(k)-(1-k^2){\rm \bf  K}(k)]}{3(1+k^2)^{3/2}}\frac{v}{\sqrt{1-v^2}}.
\eea
It is easy to verify that $v$ is a velocity of the center-of-mass of the system.

4. Let us now proceed with the study of quantum correction to the ground state energy, for which the equation
\bea
(H_2-E_2)\Psi_0=0
\eea
has to be solved.
The operator $H_2$ is quadratic form of operators  $\Phi(x)$ and $\Pi(x)$ and can therefore be  diagonalized and reduced to an infinite set of oscillators. The only problem is to take into consideration the constraints that are imposed on $\Phi(x)$ and $\Pi(x)$. From now on we can work in the center-of-mass system by setting $v=0$, or $I=0$. This simplifies the expression for $H_2$, indeed:
\bea
H_2=\int_0^Ldx\left\{\frac{1}{2}\Pi^2(x)+\frac{1}{2}\left(\frac{\partial\Phi(x)}{\partial x}\right)^2+\mu^2(3\phi_0^2(x)-1)\Phi^2(x)\right\},
\eea
Let the functions $V_n(x)$ be an orthonormal  set of solutions of the equation
\bea
\left\{-\frac{d^2}{dx^2}+\mu^2(\frac{6k^2}{1+k^2}{\rm \bf {sn}}^2(\sqrt{\frac{2}{1+k^2},k}\mu x)-1)\right\}V_n(x)=\mathcal{E}_n^2V_n(x).
\eea
We now consider the expansions of operators $\Phi(x)$ and $\Pi(x)$ in terms of $V_n(x)$:
\bea
\Phi(x)=\sum_n{'}\sqrt\frac{1}{2\mathcal{E}_n}\left[a_nV_n(x)+a_n^{+}V_n^{*}(x)\right], \\
\Pi(x)=i\sum_n{'}\sqrt\frac{\mathcal{E}_n}{2}\left[a_n^{+}V_n^{*}(x)+a_nV_n(x)\right],
\eea
in which the prime denotes that the sum does not contain the modes with zero energy (translational mode).
This system of functions fulfills the constraint
\bea
\int_0^LdxN(x)V_n(x)=0.
\eea
We may choose without loss of generality $N(x)=m\frac{\partial\phi_0(x)}{\partial x}$.
Furthermore, the condition of completeness of the functions $V_n(x)$ is:
\bea
\sum_n\left[V_n(x)V_n^{*}(y)+V_n(y)V_n^{*}(x) \right]=A(x,y)
\eea
which indicates, that this system of functions does not contain the function corresponding to the zero mode.

The creation and annihilation operators $a^{+}_n$ and $a_n$ obey the commutation relation
\bea
[a_n,a_n^{+}]=1.
\eea
 After some notations, equation (37) can be reduced to the following form.
 \bea
 \frac{d^2V_n(z)}{dz^2}+[\lambda+N(N+1)k^2{\rm \bf {sn}}^2(z,k)]V_n(z)=0,
 \eea
 which is a Lam\'{e} equation. The notations we have introduced are:
 \bea
 z=\sqrt{\frac{2}{1+k^2}}\mu x, \quad \lambda=\frac{(\mathcal{E}_n^2+2\mu^2)\sqrt{1+k^2}}{\sqrt{2}}
 \eea
 In our case of double-well potential N=2.  Periodic solutions of the Lam\'{e}
 equation are well studied\cite{arsc}. All the eigenvalues of the Lam\'{e} equation are discrete and corresponding solutions are called Lam\'{e} polynomials. There are in general
 $2N+1$ discrete eigenvalues for given $N$ and $6$ in our case, one of them is zero, which is excluded from
 the set of functions $V_n(x)$. This is also confirmed by the completeness condition of  the functions $V_n(x)$. It is important, that among these eigenvalues only one is negative (let it denote by
 $\mathcal{E}_{-1}$), namely
 \bea
 \mathcal{E}_{-1}^2=2\mu^2(1-2\frac{\sqrt{1-k^2(1-k^2)}}{1+k^2})<0.
 \eea
All other eigenvalues are positive. On substituting the expansions (38) and (39) into (36) we obtain the diagonal
form  of $H_2$:
\bea
H_2=\frac{1}{2}{\sum_n}{'}\mathcal{E}_n[a^{+}_na_n+a_na^{+}].
\eea
The energy of the lowest state (we name it quantum periodic instanton) $E_2=\frac{1}{2}{\sum_n}{'}\mathcal{E}_n$ diverges, but this is beyond the scope of our interest, since we want to estimate the lifetime of our physical system  (one can find the details of regularization in\cite{paw1}).
Obviously, the energy of the system is a complex quantity.
The imaginary part of the energy $\Im{E}=\frac{\Gamma}{2}$ (with $\Gamma $ being a decay width)
is a measure for the lifetime $t_l$ of the system:
\bea
t_l=\frac{1}{2\Im{E}}=\frac{1}{2\mu\sqrt{2(2\frac{\sqrt{1-k^2(1-k^2)}}{1+k^2}-1)}}
\eea
In the limit $k\rightarrow1$ ($L\rightarrow\infty$) the lifetime $t_l$ approaches the $\infty$, as it should be for a stable field configuration. In the opposite limit of $k\rightarrow0$ the lifetime $t_l=\frac{\sqrt{2}}{\mu}$ is a finite quantity. This is the lifetime of a sphaleron, a configuration at the top of the potential barrier.

5. Let us discuss obtained results. First of all note, that the classical equation for the field $\phi(t,x)$ is invariant under Lorentz transformation. Using translational invariance in time and space makes it possible to consider the field as $\phi(x-vt)$ such that the equation reduces to
\bea
-(1-v^2)\frac{d^2}{d\phi(z)^2}  +U'(\phi(z))=0
\eea
with $z=x-vt$, which is exactly the equation (27).
Nevertheless we have seen that the Lorentz invariance has been violated. We are convinced that this invariance is restored
when $k=1$. Indeed the solution (28) becomes:
\bea
\phi(x)=\tanh(\frac{\mu x}{\sqrt{1-v^2}}),
\eea
which describes a moving kink. The period $L$ tends to the  infinity (a line).
The energy and the momentum take the the form:
\bea
E_0=\frac{8\mu n}{3\sqrt{1-v^2}}, \qquad I=\frac{8\mu n v}{3\sqrt{1-v^2}}.
\eea
This quantities can be rewritten as
\bea
E_0=2nE_k, \qquad  I=2nI_k
\eea
with $E_k=4\mu n/3\sqrt{1-v^2}$ and $I_k=4\mu n v/3\sqrt{1-v^2}$ being the energy and the momentum
of a kink. Thus in the limit of an infinite line the solution (28) describes a kink, while the energy and the
momentum correspond to the system of $2n$ kink-antikink. The Lorentz invariance is evident.

When $k=0$ the field $\phi(x)=0$ and matches the top of the potential barrier - this is a sphaleron\footnote{We prefer to retain the term "periodic instanton" for periodic field configurations, that is often used in the literature\cite{mue1} and
references there)}.The period $L=L_n=\sqrt{2}\pi n\sqrt{1-v^2}/\mu $. One need to emphasize that the circumference $L$ is
subjected to the contraction: the larger the velocity $v$ the smaller $L$.  The momentum $I$ for $k=0$ is zero. Consistency of the equation of motion with periodic boundary conditions requires velocity to be zero - $v=0$. This means that the energy in (33) becomes $E_0 = \frac{\sqrt{2}\mu\pi n}{2}$.

Let us now approximate the energy and the momentum for two  limiting cases of $k$, namely for a) $k\ll1$ and b)$k\lesssim 1$ or  equivalently $k'=\sqrt{1-k^2}\ll1$. The standard expansions of complete elliptic functions in $k$ and $k'$ allows us to approximate  the energy and the momentum for any fixed $v$ as follows:
\bea
&&E_0=\frac{\sqrt{2}\mu\pi n}{2}\sqrt{1-v^2}+\frac{\sqrt{2}\mu\pi n}{8\sqrt{1-v^2}}[3k^2-\frac{87}{16}k^4+v^2(13k^2-\frac{329}{16}k^4)],\\
&&I=\frac{2\sqrt{2}\mu n \pi v}{\sqrt{1-v^2}}\left(k^2-\frac{13}{8}k^4\right), \\
&&k\ll1 \nn.
\eea
For  completeness we give for $k\ll1$ the expressions of the solution (28) and the lifetime:
\bea
&&\phi_0(x)=k\sqrt{2}sin\frac{\sqrt{2}\mu x}{\sqrt{1-v^2}},  \\
&&t_l=\frac{\sqrt{2}}{4\mu}+\frac{3\sqrt{2}}{8\mu}\sqrt{1-v^2}(k^2+k^4).
\eea
It is also interesting  to rewrite these quantities in terms of circumference $L$. These are:
\bea
E_0&=&\frac{\sqrt{2}\mu n \pi}{2}\sqrt{1-v^2}+\frac{\sqrt{2}\mu n \pi}{2\sqrt{1-v^2}}\left\{\frac{L-L_n}{L_n} -\frac{87}{36}\frac{(L-L_n)^2}{L_n^2}+ \right. \\ \nn
&& \left.v^2\left[\frac{13}{3}\frac{L-L_n}{L_n}-\frac{329}{36}\frac{(L-L_n)^2}{L_n^2} \right]\right\},\\
I&=&\frac{8\sqrt{2}\mu\pi nv}{3\sqrt{1-v^2}}\left\{\frac{L-L_n}{L_n}-\frac{13}{6}\frac{(L-L_n)^2}{L_n^2}\right\},  \\  t_l&=&\frac{\sqrt{2}}{4\mu}+\frac{\sqrt{2}}{2\mu}\left\{\frac{L-L_n}{L_n}+\frac{4}{3}\frac{(L-L_n)^2}{L_n^2}\right\}.
\eea
The expression for the energy at $v=0$ differs from that one obtained in\cite{sam}, namely the coefficient of the second
term in (56) is $87/36$ whereas in the cited paper is $8/3$. This difference can be explained by assumption made in\cite{sam},
 namely because of periodicity of $\phi_0(x)$ the derivative  $\partial \phi_0(x)/\partial x$ has been chosen to be zero
 at $x=0$ and $L$. We did not make such assumption. Correspondingly, they contribute to the expression for energy.
 It should be noted that the coincidence of the coefficients of the first term takes place due to the fact that when $L=L_n$
 the energy (56) must correspond to the sphaleron. This happens at $L=L_n$ (and $v=0$). One sees that the formulae (56)-(58)
 correspond to those ones at  $L=L_n$  (and $v=0$).

 We now turn to the opposite limit  $k\rightarrow 1$ or $k'=\sqrt{1-k^2}\rightarrow 0$. Using the expansion of $\rm {\rm \bf  K}(k)$
 one gets for $k'$:
 \bea
 k'^2=16\exp(-\frac{L\mu}{2n\sqrt{1-v^2}})  \nn
 \eea
 expanding the energy, the momentum and the lifetime in $k'$ up to $k'^4$ gives the following results:
 \bea
&&E_0=\frac{8\mu n}{3\sqrt{1-v^2}}-\frac{32 \mu n}{\sqrt{1-v^2}}\left(1+\frac{v^2L\mu}{n\sqrt{1-v^2}}\right)\exp(-\frac{L\mu}{n\sqrt{1-v^2}}), \\                                                            &&I=\frac{8\mu nv}{3\sqrt{1-v^2}}-\frac{32 \mu nv}{\sqrt{1-v^2}}\exp(-\frac{L\mu}{n\sqrt{1-v^2}}), \\
 &&t_l=\frac{\sqrt{3}}{48\mu}\left(1-8\exp(-\frac{L\mu}{2n\sqrt{1-v^2}})\right)\exp(\frac{L\mu}{2n\sqrt{1-v^2}}).
\eea
These formulae describe the system near the vacuum. One sees again that the Lorentz relation between the energy and the  momentum holds for $L\rightarrow \infty$. Furthermore the energy (59) receives  the additional term as compared to the formula (15b) in\cite{sam}, namely second term, that describes the interaction energy between kink and antikink in addition to the factor $\sqrt{1-v^2}$ now depends on velocity $v$ (in parentheses in front of the exponential).  In a similar manner in formula (56) additional term appears that depends on $v$. One could rewrite these expansions
for $L\gtrsim L_n$ and $L\gg L_n$ in the nonrelativistic limit, but we do not give the corresponding expressions since they are too long. We only mention that there is no known relations between the energy and the momentum of the system. They will be restored in the infinite line limit. Namely:

\bea
&&E_0=\frac{8n\mu}{3}+\frac{8n\mu}{3}\frac{v^2}{2}, \\
&&I=\frac{8n\mu}{3}v, \\
&&{\rm for}\; L\rightarrow\infty \; {\rm and}\; v\ll 1.  \nn
\eea

The lifetime is evaluated for the center-of-mass of the system. The formula (58) gives the lifetime  near the  top of the potential barrier and in case $L=L_n$ one obtains a finite value. In the opposite limit $L\rightarrow \infty$ the dominant term in (61) is the exponential $\exp(\frac{L\mu}{2n\sqrt{1-v^2}})$, which is growing up infinitely as $L\rightarrow \infty$.

\section*{Acknowledgments}
This article is carried out under the project supported by the Grant of the Georgian National Science Foundation GNSF/ST-08/4-405.

\end{document}